\documentclass[10pt,letterpaper]{article}
\usepackage{cite}
\usepackage{opex3}
\usepackage{geometry, color, graphicx}
\usepackage{mathptmx, courier, helvet} 
\usepackage{amsmath,amssymb}
\begin{document}

\title{Path length enhancement in disordered media for increased absorption}

\author{Rajeshkumar Mupparapu,$^{1,2}$ Kevin Vynck,$^{1,3}$ Tomas Svensson,$^{1}$ Matteo Burresi,$^{1,4}$ and Diederik S. Wiersma$^{1,5,*}$}

\address{
$^1$ European Laboratory for Non-linear Spectroscopy (LENS), Via Nello Carrara 1, 50019 Sesto Fiorentino, Italy\\
$^2$ Istituto Italiano di Tecnologia, via Morego 30, 16163 Genova, Italy \\
$^3$ Laboratoire Photonique, Num\'erique et Nanosciences (LP2N), UMR 5298, CNRS - IOGS - Univ. Bordeaux, 33400 Talence, France\\
$^4$ Istituto Nazionale di Ottica, Consiglio Nazionale delle Ricerche, Largo Fermi 6, 50125 Firenze, Italy\\
$^5$ Universit\'a di Firenze, Dipartimento di Fisica e Astronomia, via Giovanni Sansone 1, 50019 Sesto Fiorentino, Italy}
\email{$^*$ wiersma@lens.unifi.it}

%%%%%%%%%%%%%%%%% Abstract %%%%%%%%%%%%%%%%%%%%%%%%%%%%%%%%%%

\begin{abstract}
We theoretically and numerically investigate the capability of disordered media to enhance the optical path length in dielectric slabs and augment their light absorption efficiency due to scattering. We first perform a series of Monte Carlo simulations of random walks to determine the path length distribution in weakly to strongly (single to multiple) scattering, non-absorbing dielectric slabs under normally incident light and derive analytical expressions for the path length enhancement in these two limits. Quite interestingly, while multiple scattering is expected to produce long optical paths, we find that media containing a vanishingly small amount of scatterers can still provide high path length enhancements due to the very long trajectories sustained by total internal reflection at the slab interfaces. The path length distributions are then used to calculate the light absorption efficiency of media with varying absorption coefficients. We find that maximum absorption enhancement is obtained at an optimal scattering strength, in-between the single-scattering and the diffusive (strong multiple-scattering) regimes. This study can guide experimentalists towards more efficient and potentially low-cost solutions in photovoltaic technologies.
\end{abstract}


%%%%%%%%%%%%%%%%%%%%%%% References %%%%%%%%%%%%%%%%%%%%%%%%%

%%%%%%%%%%%%%%%%%%%%%%%%%%%  body  %%%%%%%%%%%%%%%%%%%%%%%%%%
\section{Introduction}
\begin{figure}[ht!]
\begin{center}
\includegraphics[width=0.8\textwidth]{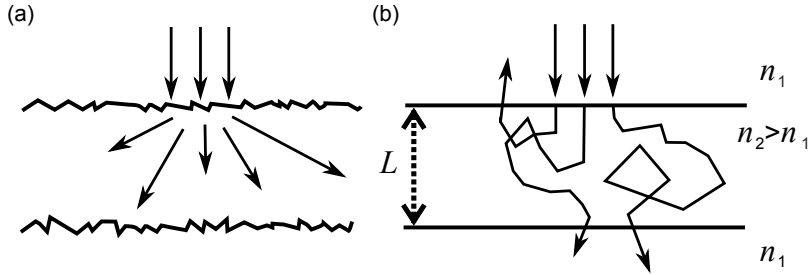}
\caption{Illustration of possible approaches to enhance light absorption in dielectric slabs. (a) Random roughness is introduced on the surfaces of the slab, leading to an efficient spreading of light in the medium. At broad angles, light follows long trajectories. (b) Volume scattering is exploited to create a broad distribution of path lengths in the medium, consisting of trajectories that may be smaller or much larger than the sample thickness.}
\label{fig:1}
\end{center}
\end{figure}

Maximizing the light absorption efficiency of dielectric media is paramount especially in photovoltaic technologies. In general, enhanced absorption is provided by engineering the surface and/or the volume of the absorbing medium as to increase the time that light interacts with it. While attention has mainly been driven towards thin (wavelength-scale) films in recent years, where strong coherent effects from photonic and/or plasmonic nanostructures can be exploited~\cite{Atwater_Natmater2010, Mallick_MRS2011, Green_Natphotonics2012, Priolo_Natnanotech2014}, thicker films, such as the standard silicon photovoltaic cell, luminescent solar concentrators~\cite{Goetzberger1977,Weber1976}, or dye-sensitized solar cells~\cite{Gratzel2003}, are widely used in practice. The standard strategy to enhance the absorption efficiency of thick dielectric films consists in designing structured surfaces (e.g., gratings, random roughness, scattering particles, etc.) as to reduce the reflection caused by the refractive index contrast at the interface, and increase the optical path length in the medium by enabling an efficient light coupling to long trajectories in the medium~\cite{yablonovitch_intensity_1982, Green_progressinphotovoltaics2002, Redfield_APL1974, Heine_AO1995, Sprafke_wiley2015}, see Fig.~\ref{fig:1}(a). A viable alternative to provide longer optical path lengths is to exploit the volume scattering provided by disordered media~\cite{Rothenberger1999, usami_theoretical_2000, Muskens2008, Mupparapu_2012}, see Fig.~\ref{fig:1}(b). It is well known that multiple light scattering leads to longer optical paths -- this property has been exploited, for instance, to enhance the interaction of light with gas in porous media for spectroscopy purposes~\cite{Svensson2011} -- yet short optical paths, which dominate in reflection, tend to reduce the overall absorption of the medium. The respective contribution of short and long optical paths evidently depends on the absorption strength of the medium, suggesting that an optimal scattering strength, obtained by tuning the size and/or density of the scattering particles, should be used to reach maximum absorption enhancement. This optimum has recently been observed in numerical simulations on dye-sensitized solar cells~\cite{Galvez2014}. To date, however, the delicate and fundamental relation between the full path length distributions, including short and long trajectories, and the light absorption enhancement in weakly to strongly scattering media has not been investigated thoroughly.

In this paper, we theoretically and numerically investigate the capability of disordered media to enhance the optical path length in dielectric media, from the weak to the strong scattering regime, and maximize the light absorption efficiency. We perform a series of Monte Carlo simulations of random walks to determine the path length distribution in disordered media and derive analytical expressions for the path length enhancement in these two limits. An interesting result is the fact that large path length enhancements are obtained in both weakly and strongly scattering limits. Because the path length distributions are very different in these two cases, however, the light absorption efficiency is found to be markedly different. Considering media with varying absorption coefficients, we find that maximum absorption enhancement is obtained at an optimal scattering strength, in-between the single-scattering and the diffusive (strong multiple-scattering) regimes.

\section{Path length enhancement}

We consider a dielectric slab, infinite in lateral directions, with effective refractive index $n_2$ and thickness $L$ embedded in an environment with refractive index $n_1$. The disordered medium composing the slab is assumed to be statistically homogeneous and isotropic, such that ballistic light is attenuated following the Beer-Lambert law over a typical distance that is the scattering mean free path $\ell_s$. For simplicity, in the analytical derivations below, single scattering is assumed to be isotropic, making the so-called transport mean free path $\ell_t$ equal to $\ell_s$. Furthermore, light is assumed to be at normal incidence on the slab and completely unpolarized, making it possible to use average Fresnel coefficients for the reflection from the slab interfaces.

\subsection{Monte Carlo simulations}

In order to get insight into the potential of volume scattering for path length enhancement in dielectric slabs, we start by performing a series of Monte Carlo (MC) simulations of random walks in scattering media~\cite{Wang_1995}. This standard numerical approach is equivalent to modeling light transport by the radiative transfer equation~\cite{Chandrasekhar_book}. It is typically valid under the assumption that the scattering mean free path $\ell_s$ is much larger than the wavelength in the medium $\lambda$ ($k \ell_s \gg 1$ with $k=2\pi/\lambda$) to avoid the occurrence of mesoscopic phenomena like localization~\cite{Lagendijk2009}, which is the case for most standard disordered media. For a sufficiently large number of simulated random walker trajectories, the density of random walkers at point $\mathbf{r}$ and time $t$ is expected to be directly proportional to the light energy density averaged over disorder realizations at $\mathbf{r}$ and $t$.

The step lengths were randomly chosen from an exponential distribution, describing the Beer-Lambert law attenuation, with mean $\ell_s$, and the new direction of propagation after a scattering event was randomly and uniformly chosen on the $4\pi$ solid angle. Reflection at the slab interfaces was implemented by calculating the Fresnel coefficients depending on the angle of incidence and averaging the reflectance over polarization. Thus, when the angle of incidence was greater than the critical angle, random walkers would experience total internal reflection.

\begin{figure}[ht!]
\begin{center}
\includegraphics[width=0.9\textwidth]{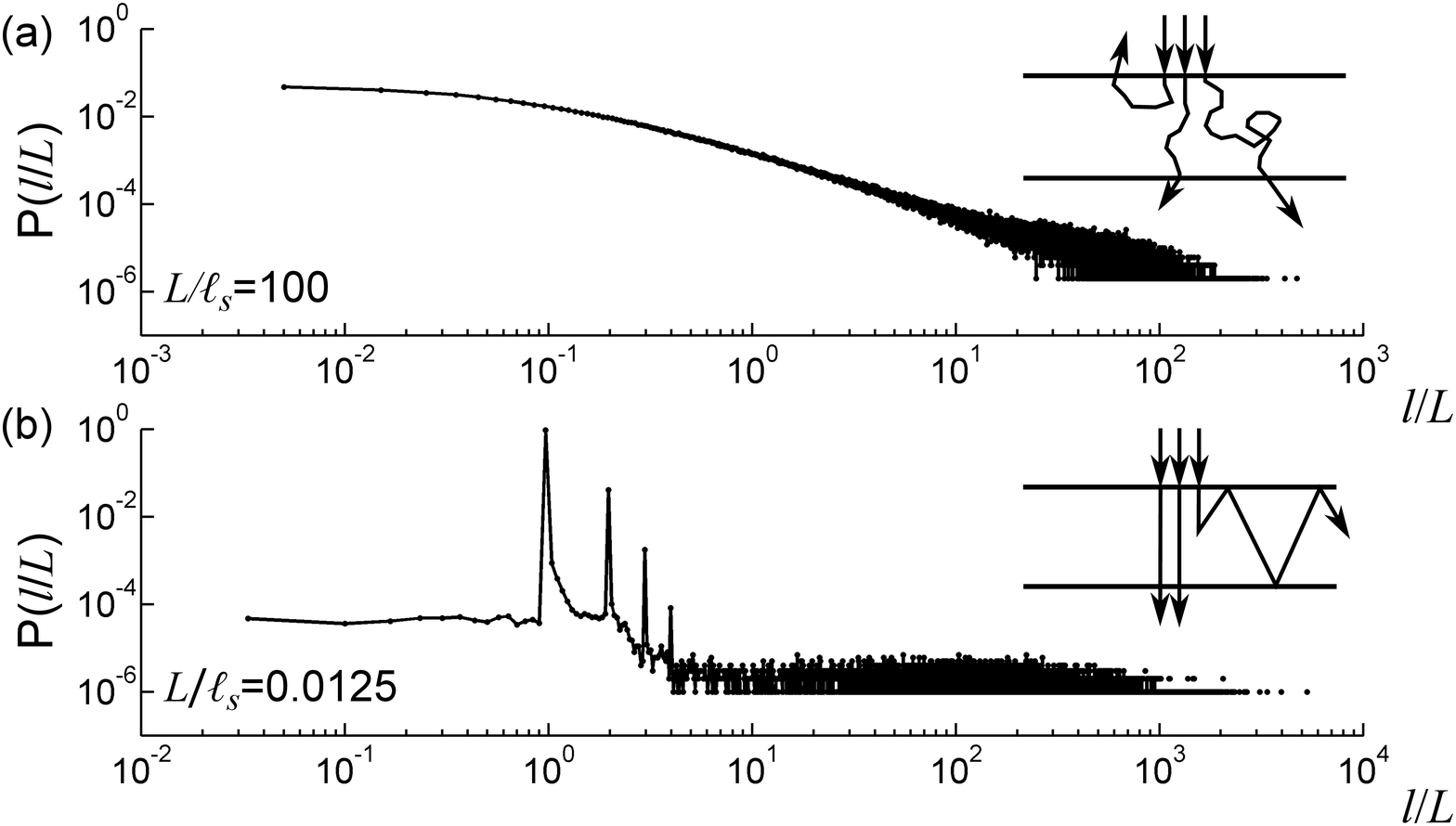}
\caption{Path length distributions $P(l/L)$ (normalized to the sample thickness) in media with different optical thicknesses. (a) Strong multiple scattering in optically thick media (here, $L/\ell_s=100$) leads to a broad and smooth distribution of path lengths. Some trajectories are shorter than the sample thickness, and others are much longer, see the inset. (b) In optically thin media (here, $L/\ell_s=0.0125$), most of the light leaves the slab without being scattered, but the small portion of scattered light can perform very long trajectories thanks to total internal reflection, see the inset.}
\label{plds}
\end{center}
\end{figure}

MC simulations were performed for a medium with $n_2=1.525$ embedded in air ($n_1=1$) and for varying optical thicknesses $L/\ell_s$, covering about 6 orders of magnitude (note that optical thickness generally refers to the quantity $L/\ell_t$; here $\ell_t=\ell_s$). This system was chosen in view of future proof-of-principle experiments, which could be performed on polymer media containing dielectric (e.g. titania) nanoparticles and dyes, since all quantities could be very well controlled (solutions of monodisperse titania particles are readily available and often used in scattering experiments, as in Ref.~\cite{SvenssonOL2013}). We simulated more than $10^6$ random walk trajectories in each case and recorded the total path length $l$ of each trajectory. Only random walkers that entered the medium were considered (those reflected by the first interface have a path length of exactly 0). Figure~\ref{plds} shows the path length distributions $P(l)$ -- the integral of which equals 1 -- in optically thick ($L/\ell_s=100$) and thin ($L/\ell_s=0.0125$) media, which are found to be markedly different. In the optically thick medium, the path length distribution is broad and smooth. This is a characteristic of efficient multiple light scattering, which creates trajectories with many different path lengths, from trajectories much shorter than the sample thickness $L$ (hence, random walkers necessarily escape from the illuminated interface) to very long trajectories, see the inset of Fig.~\ref{plds}(a). By contrast, in the optically thin medium, the path length distribution exhibits very sharp peaks in correspondence with multiples of $L$. This shows that, as expected, most random walkers escape the medium without being scattered at all. More interestingly, simulations indicate that a few random walkers have performed extremely long trajectories, orders of magnitude longer than $L$. This possibility is provided by total internal reflection: random walkers that are scattered in angles above the critical angle experience multiple internal reflections until they are scattered again, after an average distance $\ell_s \gg L$, see the inset of Fig.~\ref{plds}(b). As we will show below, these rare events are those that contribute the most to the path length enhancement.

\subsection{Theory}

We now turn to the derivation of analytical expressions for the average path length in slabs of scattering media, starting with the case of an optically thick medium, which has been investigated previously (e.g.~\cite{Arridge_PMB1992}). When the sample thickness exceeds several transport mean free paths (typically $L > 8\ell_t$~\cite{Elaloufi_JOSAA2004}), the ensemble-averaged energy density $u(\mathbf{r},t)$ can be well described by a standard diffusion equation
\begin{equation}
\frac{\partial u(\mathbf{r},t)}{\partial t} = D \nabla^2 u(\mathbf{r},t),
\end{equation}
where $D=v \ell_t/3$ is the diffusion constant, and $v=c/n_2$ is the energy velocity in the medium. In a bounded medium, $u(\mathbf{r},t)$ can be written as an eigenfunction series, the coefficients of which depend on boundary conditions. Considering that the randomization of the normally incident light takes place at a depth of one transport mean free path $\ell_t$, one can solve the diffusion equation for a point source at a depth $\ell_t$, and reach analytical expressions for the time-resolved transmission and reflection~\cite{Contini1997, Lagendijk_PhylettA1989}
\begin{eqnarray}
&& T(t) = - \frac{2 \pi D}{(L+2 z_e)^2} \sum_{n=1}^\infty n \sin \left[ n \pi \frac{\ell_t +z_e}{L+2 z_e}\right] \cos \left[ n \pi \frac{L +z_e}{L+2 z_e}\right] \exp \left[ - \frac{n^2 \pi^2 D t}{(L+2 z_e)^2}\right], \\
&& R(t) = \frac{2 \pi D}{(L+2 z_e)^2} \sum_{n=1}^\infty n \sin \left[ n \pi \frac{\ell_t +z_e}{L+2 z_e}\right] \cos \left[ n \pi \frac{z_e}{L+2 z_e}\right] \exp \left[ - \frac{n^2 \pi^2 D t}{(L+2 z_e)^2}\right].
\end{eqnarray}
Here, $z_e=\frac{1+r_i}{1-r_i}\frac{2}{3} \ell_t$ is the so-called extrapolation length and $r_i$ is the average internal reflection coefficient, which can be calculated from Fresnel coefficients~\cite{zhu_1991}.

The total transmission and reflection are therefore given by
\begin{eqnarray}
&& T_{tot} = \int_0^\infty T(t) dt = - \frac{2}{\pi} \sum_{n=1}^\infty \frac{1}{n} \sin \left[ n \pi \frac{\ell_t +z_e}{L+2 z_e}\right] \cos \left[ n \pi \frac{L +z_e}{L+2 z_e}\right], \\
&& R_{tot} = \int_0^\infty R(t) dt = \frac{2}{\pi} \sum_{n=1}^\infty \frac{1}{n} \sin \left[ n \pi \frac{\ell_t +z_e}{L+2 z_e}\right] \cos \left[ n \pi \frac{z_e}{L+2 z_e}\right],
\end{eqnarray}
and the mean transmitted and reflected times of flight by
\begin{eqnarray}
&& \langle t_T \rangle = \frac{\int_0^\infty t T(t) dt}{T_{tot}} = - \frac{2 (L+2 z_e)^2}{\pi^3 D T_{tot}} \sum_{n=1}^\infty \frac{1}{n^3} \sin \left[ n \pi \frac{\ell_t +z_e}{L+2 z_e}\right] \cos \left[ n \pi \frac{L +z_e}{L+2 z_e}\right], \\
&& \langle t_R \rangle = \frac{\int_0^\infty t R(t) dt}{R_{tot}} = \frac{2 (L+2 z_e)^2}{\pi^3 D R_{tot}} \sum_{n=1}^\infty \frac{1}{n^3} \sin \left[ n \pi \frac{\ell_t +z_e}{L+2 z_e}\right] \cos \left[ n \pi \frac{z_e}{L+2 z_e}\right].
\end{eqnarray}

The mean exit time $\langle t \rangle$ is a weighted sum of the mean transmitted and reflected times of flight, yielding
\begin{align}
\langle t \rangle &= \langle t_T \rangle T_{tot} + \langle t_R \rangle R_{tot} \\
&= \frac{2 (L+2 z_e)^2}{\pi^3 D} \sum_{n=1}^\infty \frac{1}{n^3} \sin \left[ n \pi \frac{\ell_t +z_e}{L+2 z_e}\right] \left( \cos \left[ n \pi \frac{z_e}{L+2 z_e} \right] - \cos \left[ n \pi \frac{L +z_e}{L+2 z_e}\right] \right).
\end{align}
To the lowest order in $\ell_t/L$ (i.e. for optically thick media) and performing the sum over $n$, the expression of $\langle t \rangle$ reduces to
\begin{equation}
\langle t \rangle \approx \frac{2 L}{\pi^2 D} \sum_{n=1}^\infty \frac{1}{n^2} (\ell_t+z_e) (1-(-1)^n) = \frac{L (\ell_t +z_e)}{2 D}.
\end{equation}
Finally, writing $\langle t \rangle$ explicitly in terms of $\ell_t$, a simple analytical expression for the average path length $\langle l \rangle=v \langle t \rangle$ is found
\begin{equation}\label{eq:pathlengththick}
\langle l \rangle = \left (\frac{3}{2} + \frac{1+r_i}{1-r_i}\right ) L.
\end{equation}

In the case where the refractive indices of the scattering and embedding media match ($n_1=n_2$), internal reflection is suppressed ($r_i=0$), yielding an average path length $\langle l \rangle=2.5 L$. For a medium with refractive index $n_2=1.525$ in air ($n_1=1.0$), we obtain $r_i=0.5841$ and an average path length $\langle l \rangle=5.31 L$, which is quite significant considering the low refractive index contrast. Finally, it is interesting to note that the average path length is actually independent of $\ell_t$ (or $\ell_s$). This point will be discussed lateron.

Let us now consider the case of an optically thin medium, $L/\ell_t \ll 1$. For simplicity in the analytical treatment, we assume isotropic scattering, $\ell_t=\ell_s$. As suggested by the MC simulations, only a tiny portion of the incident light is scattered in the medium, yet this leads to extremely long trajectories in the slab due to total internal reflection, which can potentially result in significant path length enhancements.

The average path length performed in the medium can be described analytically using a simple iterative scheme, that describes successive light scattering and outcoupling events
\begin{eqnarray}\label{eq:iterative}
\langle l \rangle &=& N_{i,b} \ell_{i,b} + (1-N_{i,b}) \ell_{i,s} \nonumber \\
&+& (1-N_{i,b}) P_{out} \ell_{out} + (1-N_{i,b}) (1-P_{out}) \ell_{in} \nonumber \\
&+& (1-N_{i,b}) (1-P_{out}) P_{out} \ell_{out} + (1-N_{i,b}) (1-P_{out})^2 \ell_{in} \nonumber \\
&+& ... \nonumber \\
&+& (1-N_{i,b}) (1-P_{out})^{n-1} P_{out} \ell_{out} + (1-N_{i,b}) (1-P_{out})^n \ell_{in}.
\end{eqnarray}
Here, the first line describes the light that enters the system at normal incidence, the second line the light that has been scattered once, the third line the light that has been scattered twice, etc. $N_{i,b}$ is the portion of incident light that leaves the system without being scattered (but that can be reflected on the medium interfaces one or more times) and $\ell_{i,b}$ the average distance made by it. The portion of light $1-N_{i,b}$ performs an average distance $\ell_{i,s}$ in the medium until the first scattering event. $P_{out}$ is the portion of light that may couple out from the slab after a scattering event because its propagation angle lies below the critical angle for total internal reflection. $\ell_{out}$ is the average distance made by it, from the scattering event until it is either scattered or couples out from the slab. The portion of scattered light $1-P_{out}$, instead, is guided by total internal reflection and performs an average distance $\ell_{in}$ until the next scattering event. The sum over $n$ scattering events can be simplified by the power series $\sum_{n=0}^\infty x^n=(1-x)^{-1}$, yielding
\begin{equation}\label{eq:ltot1}
\langle l \rangle = N_{i,b} \ell_{i,b} + (1-N_{i,b}) \ell_{i,s} + (1-N_{i,b}) \ell_{out} + (1-N_{i,b}) \frac{1-P_{out}}{P_{out}} \ell_{in}.
\end{equation}

Let us now calculate the various quantities in Eq.~(\ref{eq:ltot1}). The attenuation of the intensity along the light trajectory is due, on one hand, to scattering and, on the other hand, to outcoupling at slab interfaces. The attenuation of the intensity (normalized to the incident intensity) due to scattering only follows Beer-Lambert law
\begin{equation}
\tilde{I}_{s}(l) = \exp[-l/\ell_s].
\end{equation}
Regarding the attenuation due to outcoupling at slab interfaces, we consider that light propagates at an angle $\theta$ from the normal direction. At each interface, a portion of light may couple out, depending on the reflectance $r(\theta)$, calculated from Fresnel coefficients in the case of unpolarized light. Thus, as light propagates in the waveguide, it is attenuated by a factor $r$ along the path length with a period $L/\cos(\theta)$. To simplify the task, we assume that the decay of the intensity due to outcoupling may be approximated well by a simple decaying exponential
\begin{equation}\label{eq:exp_decay}
\tilde{I}_o(l)=\exp \left[- \frac{l (1- r(\theta)) \cos(\theta)}{L} \right].
\end{equation}
This approximation is expected to be valid to describe quantities involving many reflections on the interface.

At normal incidence ($\theta=0$) and writing $r(\theta=0)=r_0$, the ratio of light that couples out of the slab without being scattered is then
\begin{equation}
N_{i,b}=\frac{\int \tilde{I}_{s}(l) \tilde{I}_o(l) dl}{\int \tilde{I}_o(l) dl} = \frac{\ell_s(1-r_0)}{L+\ell_s(1-r_0)},
\end{equation}
and the average distance made it
\begin{equation}
\ell_{i,b} = \frac{\int l \tilde{I}_{s}(l) \tilde{I}_o(l) dl}{\int \tilde{I}_{s}(l) \tilde{I}_o(l) dl} = \frac{L \ell_s}{L+\ell_s(1-r_0)}.
\end{equation}
The integrals in the denominator are present for normalization purposes (the decaying intensities are not probability density functions). We clearly see that for $L/\ell_s \ll 1$, $N_{i,b} \rightarrow 1$ and $\ell_{i,b} \rightarrow L/(1-r_0)$.

Regarding the light that is scattered in the medium, the critical angle for total internal reflection, $\theta_c$, allows distinguishing two cases:
\begin{enumerate}
\item $0\leq \theta<\theta_c$: light can couple out from the slab. In the limit where $\ell_s \gg L$, the attenuation will be essentially due to outcoupling, i.e. scattering can be neglected in the estimation of $\ell_{out}$.
\item $\theta_c\leq \theta \leq \pi/2$: light is confined to the slab by total internal reflection. The attenuation is entirely due to scattering, yielding $\ell_{in}=\ell_s$.
\end{enumerate}
The critical angle is simply defined as
\begin{equation}
\theta_c=\arcsin\left[\frac{n_1}{n_2} \right],
\end{equation}
such that the portion of scattered light that may couple out from the slab is given by
\begin{equation}
P_{out} = \frac{2}{4\pi} \int_0^{2\pi} d\phi \int_0^{\theta_c} \sin \theta d\theta = 1-\sqrt{1-\frac{n_1^2}{n_2^2}}.
\end{equation}
Here, the factor 2 comes from the fact that light can escape from both sides of the slab.

Let us now consider the limit $L/\ell_s \ll 1$ in Eq.~(\ref{eq:ltot1}). As shown above, the first term will converge to a non-zero value, $L/(1-r_0)$. In the second and third terms, $(1-N_{i,b})$ goes to 0, while both $\ell_{i,s}$ and $\ell_{out}$ are expected to be on the order of $L$. Indeed, the intensity rapidly decays due to outcoupling for realistic values of the reflection coefficient (an index contrast of 1.525, for instance, yields $r_0=0.0432$). Thus, in the former case, scattering of the incident light is more likely to occur within the first multiple reflections between interfaces, and in the latter case, the average distance made by reflection between interfaces for angles near normal (below the critical angle) is very unlikely to exceed a few sample thicknesses. As a result, the second and third terms can be neglected. Finally, the last term is also proportional to $(1-N_{i,b})$ but it is totally compensated by $\ell_{in}=\ell_s$, which is much larger than $L$.

Following these considerations, inserting the derived expressions into Eq.~(\ref{eq:ltot1}) with $r_0=(n_1-n_2)^2/(n_1+n_2)^2$, and letting $\ell_s$ tend towards infinity, we obtain an expression for the average path length
\begin{equation}\label{eq:pathlengthweakscatt}
\langle l \rangle = \frac{(n_1 + n_2)^2}{4 n_1 n_2 \left( 1-\sqrt{1-\frac{n_{1}^{2}}{n_{2}^{2}}} \right)} L.
\end{equation}
When $n_1=n_2$, $\langle l \rangle=L$, i.e. there is no enhancement of the path length. This is expected since total internal reflection is absent and cannot compensate for the tiny scattering. On the other hand, when $n_1=1.0$ and $n_2=1.525$, we obtain $\langle l \rangle=4.27L$, which is surprisingly large considering the vanishingly small scattering. Thus, we have shown here that, although scattering can be extremely weak -- the slab is essentially transparent -- the path length enhancement can be remarkably large thanks to the rare but very long trajectories sustained by total internal reflection in the slab.

It is worthy of notice that Eq.~(\ref{eq:pathlengthweakscatt}), like Eq.~(\ref{eq:pathlengththick}), is totally independent of the scattering strength of the medium. This observation has, in fact, deep physical roots. The optical path length enhancement in a medium, within the radiative transfer picture, equals the light intensity enhancement. Upon average over incident angles, it is expected that the latter be bounded by the density of states ratio between the medium and its environment, i.e. by $(n_2/n_1)^2$ in 3D (yielding the so-called Yablonovitch limit~\cite{Yablonovitch}), and be independent on microscopic details. This invariance property has recently been demonstrated theoretically and numerically for wave scattering in disordered media~\cite{Pierrat_PNAS}. In view of the above considerations, it is therefore not surprising that $\langle l \rangle$ in Eqs.~(\ref{eq:pathlengththick})-(\ref{eq:pathlengthweakscatt}) is independent of $\ell_s$ or $\ell_t$. Because illumination is not isotropic in our case, different expressions are obtained in the limits of optically thin and thick media, but we expect that both should converge to a unique expression upon average over incident angles. In particular, larger incidence angles are expected to increase the average path length in optically thin media due to the longer trajectories provided by the unscattered light and to decrease it in optically thick media due to the fact that the incident light will be scattered at shallower depths, thereby augmenting the respective contribution of short trajectories in reflection. It will be interesting to investigate this aspect in a future work.

\subsection{Comparison between Monte Carlo simulations and theory}

\begin{figure}[ht!]
\begin{center}
\includegraphics[width=0.8\textwidth]{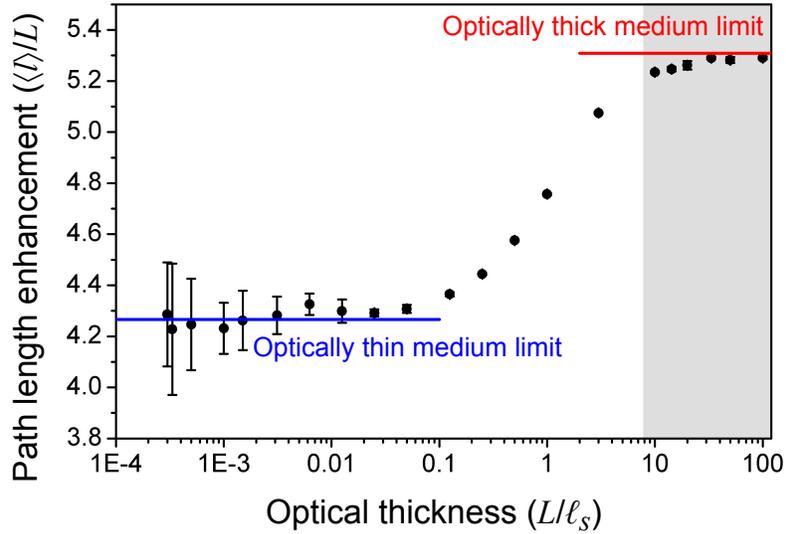}
\caption{Path length enhancement $\langle l \rangle/L$ determined from MC simulations (black dots) versus the optical thickness $L/\ell_{s}$, ranging from optically thin to optically thick media (left to right). The gray colored region indicates region where diffusion theory applies, the red and blue lines indicate the path length enhancement expected in the limit of optically thick ($L/\ell_{s} \gg 1$) and optically thin ($L/\ell_{s} \ll 1$) media. MC simulations are in very good agreement with theoretical predictions, Eqs.~(\ref{eq:pathlengththick})-(\ref{eq:pathlengthweakscatt}).}
\label{normal}
\end{center}
\end{figure}

Let us now compare the average path length obtained from MC simulations for different values of the optical thickness $L/\ell_s$ with theoretical predictions in the limits of optically thin and thick media. Monte Carlo simulations, performed with more than $10^6$ random walk trajectories for each optical thickness, have been repeated 5 times in order to provide an error on the average path length estimation. Results are shown in Fig.~\ref{normal}, where the average path length is found to increase smoothly with the optical thickness. The larger error with decreasing optical thickness comes from the fact that random walkers actually experiencing a scattering event, which contribute for most of the path length enhancement, are fewer, making the statistics poorer. Nevertheless, the agreement with theoretical predictions, Eqs.~(\ref{eq:pathlengththick})-(\ref{eq:pathlengthweakscatt}), is very good, thereby validating our theoretical model and our understanding of the origin of large path length enhancements in scattering media.

\section{Absorption enhancement}

At this stage, we can investigate the effect of volume scattering and the resulting path length enhancement on absorption. Absorption is assumed to be homogeneously distributed throughout the volume of the slab and is characterized by an absorption mean free path $\ell_i$. In the random walk picture, due to absorption, each trajectory is attenuated exponentially following Beer-Lambert law. The broad distribution of path lengths makes that different trajectories will contribute differently to the overall absorption of the medium. The medium absorption $A$ is related to the path length distribution $P(l)$, calculated previously and the integral of which equals 1, as follows
\begin{equation} \label{beers2}
A=(1-r_0) \left[ 1-\int P(l) \exp\left( \frac{-l}{\ell_i} \right) dl \right].
\end{equation}
The prefactor $(1-r_0)$ takes into account the fact that a portion of the incident light is directly reflected by the first interface before entering the medium.

The absorption of the bare slab (i.e. a uniform medium containing no scatterers) can be derived analytically from its path length distribution $P_{b}(l)$
\begin{equation}
P_{b}(l) = (1-r_0) \left[ \delta(l-L) + r_0 \delta(l-2L) + ... + r_0^n \delta(l-(n+1)L) \right],
\end{equation}
yielding a general expression for the total absorption enhancement $\Gamma$
\begin{equation} \label{beers3}
\Gamma=\frac{1-\int P(l) \exp\left(-l/\ell_i \right) dl}{\left[ 1-\exp \left( -L/\ell_i \right) \right] / \left[ 1 - r_{0} \exp \left( -L/\ell_i \right) \right]}.
\end{equation}

In weakly absorbing media, it is known that the absorption enhancement is essentially bounded by the path length enhancement. Assuming that $\ell_i \gg L$ and $\ell_i \gg \langle l \rangle$, Eq.~(\ref{beers3}) reduces to
\begin{equation}\label{eq:bound}
\Gamma=\frac{\langle l \rangle}{L}(1-r_0),
\end{equation}
as expected. Note that the actual path length in slabs containing no scatterers, excluding light reflected on the first interface, equals $L/(1-r_0)$, as shown above ($\ell_{i,b}$ tends to $L/(1-r_0)$ when $\ell_s \gg L$).

\begin{figure}
\begin{center}
{\includegraphics[width=0.9\textwidth]{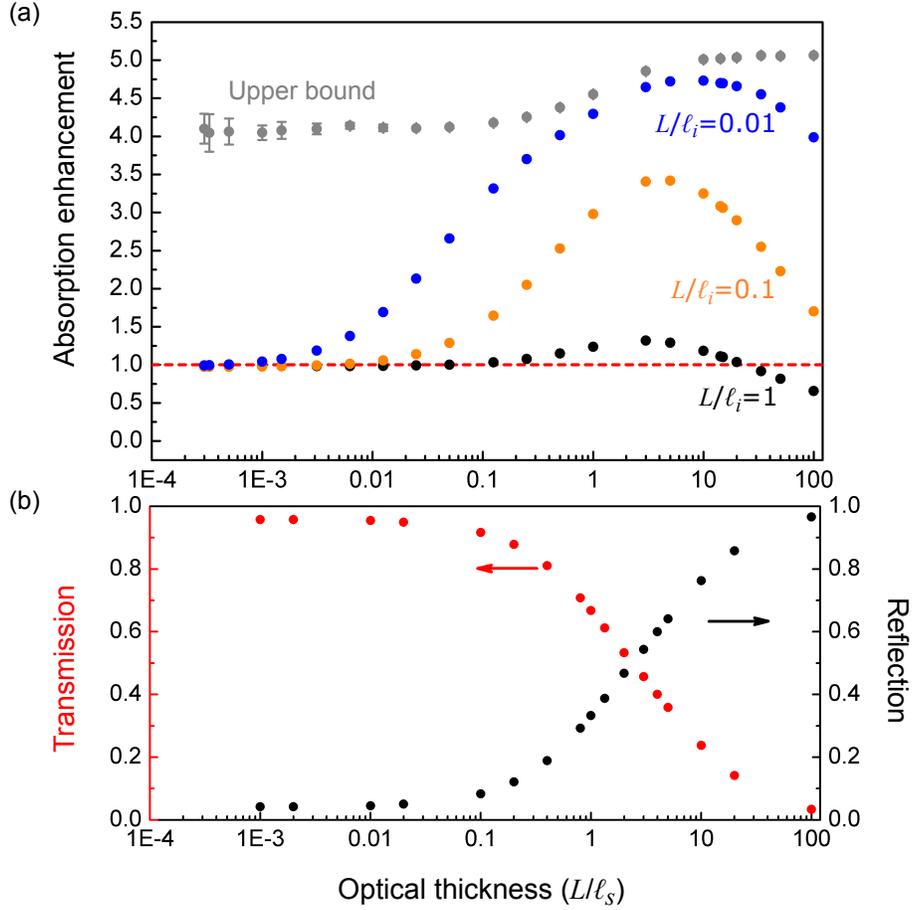}}
\caption{(a) Effect of volume scattering on light absorption in dielectric slabs for three different absorption strengths $L/\ell_i= 1, 0.1, 0.01$ (black, orange and blue dots, respectively). The gray dots indicate the upper bound for light absorption enhancement, according to Eq.~(\ref{eq:bound}). The red dashed line corresponds to media containing no scatterers (enhancement equals 1 by definition). The absorption enhancement strongly depends on the optical thickness of the medium, exhibiting an optimal regime where the enhancement is maximum. (b) Total transmission (red points) and reflection (black points) through disordered media versus their optical thickness. Transmission and reflection are comparable in an intermediate regime between single scattering and diffusion (strong multiple scattering). It is in this scattering range that light absorption enhancement is found to be maximum.}
\label{enh_absummary}
\end{center}
\end{figure}

Equation~(\ref{beers3}) was used to calculate the absorption enhancement with varying absorption strengths $(L/\ell_i= 1, 0.1, 0.01)$ by using the path length distribution retrieved previously in MC simulations on a wide range of scattering mean free paths. The enhancement factors are summarized in Fig.~\ref{enh_absummary}(a). The first observation is that the absorption enhancement is naturally higher for weakly absorbing media, approaching the upper bound in Eq.~(\ref{eq:bound}). This is expected as the absorption enhancement then benefits from the entire path length distribution and especially long trajectories, which are not immediately attenuated. More interestingly, the absorption enhancement depends strongly on the scattering strength of the medium: (i) For optically thin media, $\Gamma$ tends to remain close to 1. This is due to the fact that the rare, very long trajectories, typically longer than $\ell_s$, are more rapidly attenuated due to absorption (here, $\ell_i < \ell_s$), thereby not contributing fully to the absorption enhancement. (ii) In an intermediate regime, typically for $1 < L/\ell_s <10$, the absorption enhancement reaches a maximum. In this optimal regime, scattering is strong enough to reduce significantly the amount of light that crosses the medium without being scattered, but not too strong to allow light penetrating the medium at a sufficient depth compared to $L$. (iii) For optically thick media, although long trajectories exist and contribute to the absorption enhancement, most of the light follows trajectories that are comparable or shorter than $L$, exiting the slab from the illuminated surface, thereby preventing light from being absorbed. When both absorption and scattering are strong, this process can dominate and lead to a reduction of the overall absorption, as shown for the strongest absorbing medium in Fig.~\ref{enh_absummary}(a).

The ability of light to efficiently penetrate and explore the slab volume is illustrated in Fig.~\ref{enh_absummary}(b), where the total reflection and transmission of a non-absorbing scattering medium are plotted as a function of the optical thickness. The optimal scattering regime clearly lies in the range where the same amount of light is transmitted and reflected. With increasing absorption strength, $L/\ell_i$, the optimal scattering strength shifts towards lower values: light is more rapidly attenuated, making the presence of very long trajectories essentially unimportant.

Finally, the actual absorption of slabs, generally more relevant than the absorption enhancement for applications, is reported in Table~\ref{abstable} for different scattering and absorption strengths. For the more strongly absorbing medium ($L/\ell_i=1$), the absorption enhancement is about 30\%, leading to a total absorption of 85\%, which is rather high and competitive with approaches relying on surface engineering~\cite{Kowalczewski_OL2013}.

\begin{table}
\begin{center}
\caption{Total absorption in a bare slab and in slabs with varying optical thicknesses ($L/\ell_s$), chosen from the optimal scattering regime, and absorption strengths ($L/\ell_i$).}
\begin{tabular}{|c|c|c|c|}
\hline 
 & $L/\ell_{i}=1$ & $L/\ell_{i}=0.1$ & $L/\ell_{i}=0.01$ \\ 
\hline 
Bare slab & 0.642 & 0.099 & 0.010 \\ 
\hline
$L/\ell_{s}=1$ & 0.794 & 0.295 & 0.044 \\
\hline 
$L/\ell_{s}=3$ & 0.847 & 0.337 & 0.048 \\ 
\hline 
$L/\ell_{s}=5$ & 0.828 & 0.338 & 0.049 \\ 
\hline 
\end{tabular}
\label{abstable}
\end{center} 
\end{table}

\section{Conclusions}

In summary, we have shown that volume scattering constitutes a simple, efficient and potentially low-cost approach to enhance the optical path length in dielectric slabs and augment their light absorption efficiency. We have first demonstrated via Monte-Carlo simulations and theory that large path length enhancements can be achieved in both optically thin and thick media. In the former case, this property is provided by rare but very long trajectories sustained by total internal reflection, contrary to the latter case, where long trajectories are due to multiple scattering. We have then calculated the absorption enhancement expected in scattering media with varying optical thicknesses and absorption strengths, and revealed the presence of an optimal scattering regime, in between the single and strong multiple scattering regimes, where absorption enhancement is maximized. This optimal situation relies on an interplay between the actual absorption efficiency of the bare slab (characterized by $L/\ell_i$) and the possibility for light to efficiently explore the sample volume by scattering.

In practice, volume scattering can be obtained by embedding small scatterers (dielectric nanoparticles or even air bubbles) in the dielectric medium. Our study shows that an optimal scattering strength, driven by the scattering efficiency of the individual scatterers and the scatterer density, should be used to enhance absorption efficiency, thereby guiding experimentalists towards more efficient and potentially low-cost solutions in photovoltaic technologies. This information is particularly relevant for dye-sensitized solar cells, where the size and density of scattering titania nanoparticles can be tuned to achieve maximum light absorption without affecting their primary role in separation of charges~\cite{Rothenberger1999, usami_theoretical_2000, Galvez2014}.

Finally, let us emphasize that we have restricted our study to the case of disordered media illuminated at normal incidence, and wherein scattering is isotropic (provided by very small particles) and the scatterers are non-absorbing. It will be interesting in the future to extend this study to media composed of resonant (metallic or dielectric) particles providing anisotropic scattering and to illumination at oblique incidence. While anisotropic scattering may be taken into account by considering that the average distance after which the direction of propagating light is completely randomized is the transport mean free path $\ell_t$ (already taken into account in our diffusion model for optically thick media), oblique incidence would essentially translate into a modification of the unscattered light (first line in Eq.~(\ref{eq:iterative}) for optically thin media and position of the point source in the diffusion model for optically thick media). A further step could be to generalize this study to multilayered structures, representing real solar cells, in which case Monte Carlo simulations may be more manageable than theory, mainly due to the difficulty to deal analytically with several boundary conditions. These advances would allow us to get a complete picture of the problem at hand and propose some concrete designs of materials operating in the visible range, where absorption enhancement is greatly enhanced, and that could eventually be tested experimentally.

\section*{Acknowledgments}
We acknowledge ENI S.p.A, the Eu-NoE "Nanophotonics for Energy Efficiency" for their financial support. 

\end{document}